\documentclass[conference,10pt,letterpaper]{IEEEtran}
\usepackage{cite}
\usepackage{amsmath,amssymb}
\usepackage{graphicx}
\usepackage{booktabs,array}
\usepackage{xspace}
\usepackage{tikz}
\usetikzlibrary{arrows.meta,positioning,fit}
\usepackage[hidelinks]{hyperref}
\usepackage{xurl}
\hypersetup{pdfauthor={},pdftitle={FuseOS: Fused Persistent Execution for Dynamic Inference},pdfsubject={Anonymous IPDPS manuscript}}
\newcommand{\sys}{\textsc{GPUOS}\xspace}
\newcommand{\code}[1]{\texttt{\detokenize{#1}}}
\newcommand{\aidraft}{~\cite{Codex}}
\newcolumntype{P}[1]{>{\raggedright\arraybackslash}p{#1}}
\begin{document}
\bstctlcite{IEEEexample:BSTcontrol}
\title{GPUOS: A GPU Operating System Primitive for Transparent Operation Fusion}
\author{Yiwei Yang, Xiangyu Gao, Yuan Zhou, Yuhang Gan, Yusheng Zheng, Andi Quinn}
\maketitle
\begin{abstract}
Dynamic inference changes its live requests, ready work and transient-storage
requirements during execution. FuseOS couples these decisions in a fused
persistent resource pool: the GPU retires requests, assigns ready tiles and
reuses scratch at phase-completion boundaries. A Qwen3-0.6B integration executes
all decoder layers and remaining greedy tokens in one cooperative launch after
native prefill. Paired gate/up and activation fusion halves global temporary
storage from 24 to 12 KiB per live request, while lifetime overlay prevents
independent handler arrays from accumulating in shared memory. Across four
matched cases on an RTX 5090, fused persistent decode is 1.14--1.20$\times$
faster than the best measured tile-grid+PDL or CLC+PDL variant, allowing either
fused or separate MLP handlers. A launch-only reservation control rules out
shared-memory capacity alone as the explanation for the overlay benefit.
Fixed region planning matches per-token rebuilding, locating the useful
allocation mechanism in reuse rather than repeated planning. Real four-client
MPS comparisons show benefits and run variation for trusted request pooling.
Complete-request and ResNet-18 benchmarks expose integration limits. Matched
custom controls are bitwise identical; two of six Qwen cases pass the unchanged
native numerical gate, so native-model speedups are qualified separately.
AI assistance is disclosed in Acknowledgments~\cite{Codex}.
\end{abstract}
\begin{IEEEkeywords}
GPU runtime, dynamism, persistent kernels, load balancing, workspace allocation
\end{IEEEkeywords}
\section{Introduction}
GPU sharing becomes a dynamic scheduling problem even when every request
runs the same neural network. Requests enter with different context lengths,
produce different amounts of output, and stop at different iterations.
The remaining requests should gain access to the released execution capacity,
while their intermediate buffers must remain valid until the last consumer
finishes. We study this \emph{execution dynamism}: a changing live request set,
ready-work extent, and transient-storage demand inside an admitted batch.\aidraft{}

\begin{figure*}[t]
\centering\includegraphics[width=\textwidth]{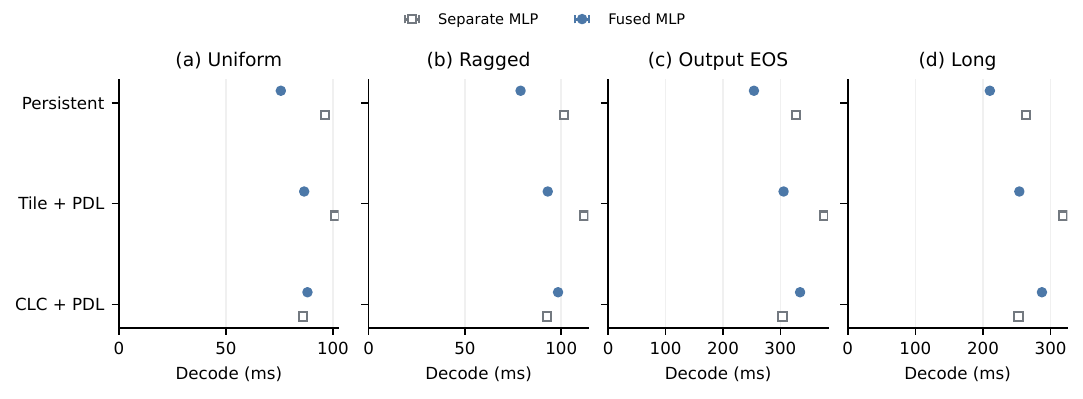}
\caption{Matched fusion $\times$ execution comparison. Markers show complete
decode latency; intervals span three process medians, each based on ten
requests. Open squares materialize gate/up and execute SiLU separately;
filled circles fuse that work. All six paths produce bitwise-identical tokens
and logits for each case. Tile and CLC grids are unrestricted by the persistent
worker count. Lower is better.}
\label{fig:factorial}
\end{figure*}

Process-level sharing and instance partitioning expose useful resource controls,
but do not identify an inference phase's ready tiles or last buffer consumer.
MPS multiplexes CUDA clients and supports resource adjustment; MIG partitions
supported GPUs into hardware instances~\cite{NVIDIAMPS,NVIDIAMIG}.
For trusted requests to one model, an alternative is to merge their ready work
into a software pool and let the GPU redistribute that work as requests finish.
This moves the sharing decision from client or instance boundaries to the
computation's own readiness and lifetime boundaries.

\sys implements this alternative with fused persistent execution. After native
prefill, one cooperative launch executes all remaining Qwen decoder layers and
greedy tokens. The GPU retires requests, compacts their identifiers, derives
ready tile bounds, and maps those tiles to a common worker pool. The same phase
boundary that releases dependent work also permits storage reuse. A request
can stop because of a token produced during execution, without a host decision
or per-token kernel launch.

The central engineering problem is that fusion can defeat its own purpose.
Combining handlers removes launches and intermediates but can retain all their
shared-memory arrays for the lifetime of the persistent kernel. In our decoder,
fusing normalization, gate/up projections, and SiLU removes a global phase,
yet separately allocated handler scratch raises the block's shared allocation
from 21.3 to 35.8 KiB. Reusing scratch according to phase lifetimes recovers the
benefit. On ragged requests, the complete decode falls from 116.0 to 78.9 ms
with identical tokens and logits. A control that changes only launch-time reservation leaves performance
unchanged, ruling out capacity alone as the cause. Fusion must therefore be
evaluated together with the generated execution path, not just its phase
count or nominal storage footprint.

Persistent kernels, request retirement and memory reuse are established
techniques. MPK already supports device-side serving decisions and shared-memory
management~\cite{Cheng2026MPK}; Meld also uses a GPU execution wrapper for
cross-SM cooperation~\cite{He2026Meld}. \sys explores a narrower design:
implicit tasks within globally ordered phases, a bounded active map, and a
common last-consumer rule for block-local and global scratch. This restriction
removes the need for per-task dependency queues in this implementation, at the
cost of global barriers and no cross-phase pipelining. The contribution is this
specialized execution design and the evidence identifying when its fusion and
storage choices produce an advantage.

We make three contributions:
\begin{itemize}
\item A persistent resource pool for finite, dynamic inference batches, with
GPU-side retirement, stable request ownership and ready-tile assignment.
\item A phase-lifetime contract that supports vertical operator fusion and
overlay of nonoverlapping scratch, reducing the demonstrated global temporary
plan from 24 to 12 KiB per live request.
\item Matched fusion and launch controls, real four-client MPS comparisons,
and complete-model benchmarks that separate pooling, persistence, allocation
and numerical compatibility.
\end{itemize}

Across four factorial cases, fused persistent decode is 1.14--1.20$\times$
faster than the best measured tile-grid+PDL or CLC+PDL variant, allowing either
fused or separate MLP handlers. A fixed region plan performs nearly identically
to rebuilding the plan every token: lifetime reuse is useful, but per-token
rebuilding is unnecessary for this bounded contract. The native-model numerical
gate passes two of six cases; matched custom controls remain bitwise identical.
These results support a systems design for trusted pooled requests, with a
specific implementation and validation boundary.

\section{Dynamism and the Sharing Boundary}
\label{sec:background}
A model's operators can be fixed while its useful work is unknown in advance.
The current token determines whether a request remains live; that decision
changes the next phase's rows, attention contexts and storage extents.\aidraft{}
Three runtime decisions follow: which work is ready, which workers may execute
it, and when its storage can be reused. Fusion is safe only when these decisions
refer to the same dependency and ownership structure.

\subsection{From Client Partitions to Ready Work}
In process-based inference, each client owns a launch stream and request state.
MPS lets those streams share GPU execution resources, including configurable
resource limits~\cite{NVIDIAMPS}. Its allocation unit does not encode the
application's last-consumer relation. A process can finish while another still
has substantial work, and the remaining process must expose enough parallelism
to use the released capacity. MPS can redistribute resources; it does not
automatically fuse the clients' projection tiles or share their temporary
storage plan.

\sys instead places compatible requests in one trusted execution domain. Dense
projection tiles cover up to four live requests, sharing weight loads. As the
active map shrinks, workers execute the remaining request tiles without
process ownership. The relevant comparison is therefore a pooled executor
against concurrently executing clients, with a pooled ordinary-kernel control
to distinguish batching from persistence. Section~\ref{sec:sharing} implements
this comparison using real MPS clients and shared model weights.

MIG addresses a different requirement: hardware-partitioned instances on
supported devices~\cite{NVIDIAMIG}. A software pool can replace an instance-based
placement policy when requests trust one another and require work conservation,
but it does not supply MIG's memory or fault isolation. The RTX 5090 reports
MIG as unavailable, so the evaluation contains no measured MIG configuration.

\subsection{Persistent Execution and Related Systems}
Persistent-thread programming has long exposed both benefits and performance
losses on GPU workloads~\cite{Gupta2012PT}. MPK compiles tensor programs to
SM-level task graphs, uses decentralized scheduling, and supports request
retirement, continuous admission and paged KV management. It selects
batch-specialized graphs and manages shared-memory pages to enable task
pipelining~\cite{Cheng2026MPK}. These capabilities precede our design. \sys
specializes to phase-ordered work: implicit row/head indices replace explicit
task dependency records, and all consumers finish before the next phase
reuses storage. No dedicated scheduler SM or shared-memory page allocator is
needed for this restricted schedule.

Meld separates computation from topology-aware cooperation plans and provides
dynamic cross-SM coordination~\cite{He2026Meld}. \sys uses cooperative grid
barriers and global intermediates; it does not implement Meld's remote
shared-memory transport or cooperation plans. Table~\ref{tab:related} states
the mechanism distinction. In MPK's public revision \code{60ae655}, a source
compatibility probe rejects CC~120 in its resource helper, and linear dispatch
excludes that architecture~\cite{MPKSource}. This is a compatibility check, not
an MPK build or timing result. Empirical MPK/Meld comparison remains open;
neither system is assigned a latency inferred from another device or model.

\begin{table}[t]
\caption{Mechanism comparison. Rows describe architectural choices, not a
performance ranking or a claim that device-side dynamism is new.}
\label{tab:related}\centering\small\setlength{\tabcolsep}{4pt}
\begin{tabular}{@{}P{13mm}P{31mm}P{35mm}@{}}\toprule
System & Scheduling boundary & Storage / dependency boundary \\\midrule
MPK~\cite{Cheng2026MPK} & SM tasks, event queues, decentralized schedulers & Paged shared memory; cross-task overlap \\
Meld~\cite{He2026Meld} & GPU wrapper and cooperation plans & Dynamic cross-SM cooperation \\
\sys & Implicit ready tiles; one active map & Global phase completion; scratch overlay \\
\bottomrule\end{tabular}
\end{table}

CUDA conditional graphs can also express device-dependent loops behind stable
pointers~\cite{CUDAConditional}. PDL relaxes launch dependencies with an explicit
consumer acquire, while CLC lets blocks claim canceled, unstarted block
indices~\cite{CUDAPDL,CUDACLC}. Our controls combine these mechanisms with the
same active map, handlers and region plan. Dynamism does not require graph
recapture, and the comparison does not rely on charging a host loop to every
alternative.

\section{Fused Persistent Resource Pool}
\label{sec:execution}
\begin{figure*}[t]
\centering
\includegraphics[width=\textwidth]{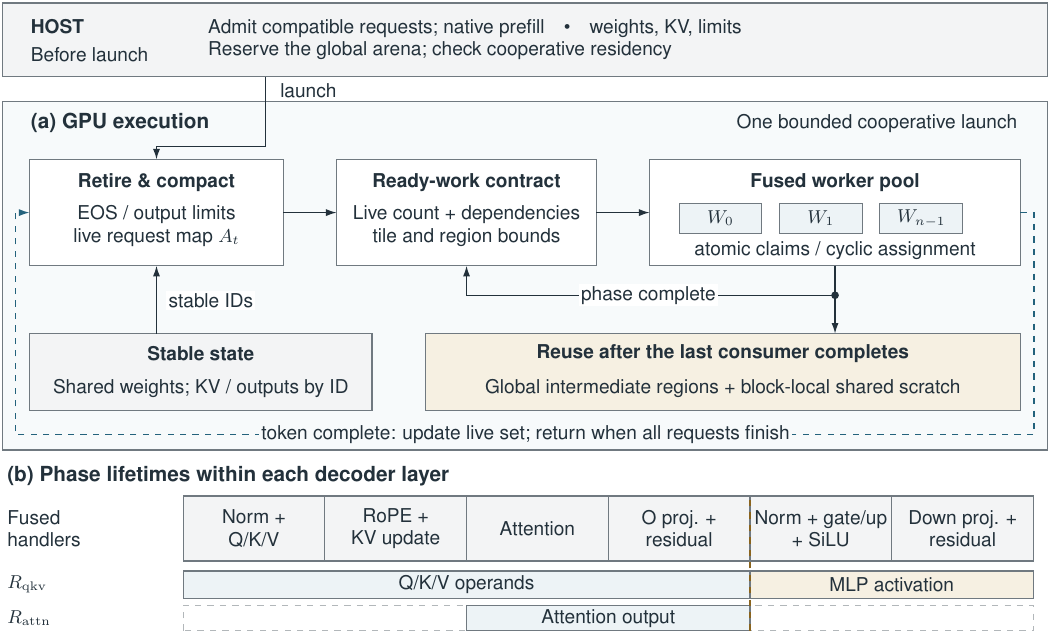}
\caption{\textbf{GPUOS execution and storage contract.}
(a) After host admission and prefill, one cooperative invocation owns the
decode loop. GPU retirement updates the live map; the phase contract derives
ready tiles for a common pool of worker blocks. Phase completion makes the
next work ready and permits scratch reuse. The dashed feedback is taken only
after a complete token. Stable IDs retain KV/output ownership independently
of packed execution positions; blocks have no fixed request or SM ownership.
(b) Six fused phases share a bounded region plan. The Q/K/V region becomes
MLP storage after its last consumer. Empty intervals indicate no live value,
not a released physical allocation; KV remains reserved until return.}
\label{fig:execution}
\end{figure*}

\sys admits a finite set of compatible requests into one execution domain.
Weights and KV buffers are persistent request state; intermediate tensors
belong to phases.\aidraft{} The host prepares inputs and performs prefill.
A cooperative kernel owns the remaining decode loop, returning when all
requests stop or reach their generation limits.
Figure~\ref{fig:execution} shows how live-work updates and scratch reuse share
this execution contract while request-owned state retains stable identities.

\begin{figure*}[t]
\centering
\includegraphics[width=\textwidth]{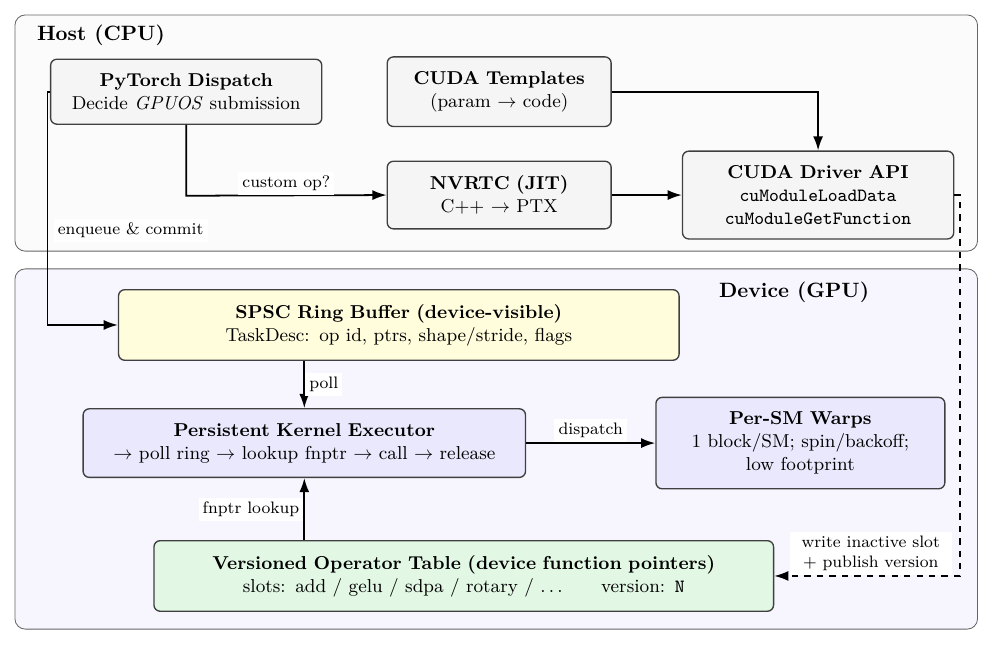}
\caption{\textbf{GPUOS system architecture.}}
\label{fig:sys_arch}
\end{figure*}

\subsection{Dynamic State and Stable Ownership}
Each request $r$ has a prompt start offset, output limit $g_r$, current token
and stable KV/output pointers. The active map $A_t$ translates packed execution
positions to stable request identifiers at step $t$. Before producing token
$t\geq1$, with the prefilled token indexed zero,
\begin{equation}
A_t=\{r\in A_{t-1}:t<g_r\ \wedge\ y_{r,t-1}\ne\mathrm{EOS}\}.
\end{equation}
A control thread compacts surviving identifiers in place. A grid barrier
publishes the map before workers derive task bounds. Retirement can therefore
depend on an output computed in the preceding iteration. Tests can disable
EOS to keep output work fixed across numerical implementations.

Intermediate rows use packed indices, whereas KV, final logits and tokens use
stable request IDs. Compaction never transfers one request's KV state to another.
Left-padded prompts share physical cache capacity but carry individual start
offsets; attention masks invalid keys and rotary positions use logical offsets.
The scan is linear in the admitted batch, capped at sixteen requests by the
implementation. It is small control work, not a general continuous-admission
queue. Retired KV allocations remain reserved until the invocation returns.

\subsection{Ready-Tile Scheduling}
The phase schedule admits only tiles whose inputs are complete and whose
outputs are disjoint. A projection task covers sixteen output rows and up to
four live requests; attention tasks cover request/head pairs. The current live
count determines valid projection tiles, and each attention task masks its own
context. No host-generated task list is required after request retirement.

In atomic mode, a block leader fetches the next index from the phase cursor,
broadcasts it to the block, and repeats after completing that tile. A cyclic
alternative assigns indices $w+kW$ to worker $w$ in a pool of $W$ blocks.
Both policies use the changing active map; cyclic assignment does not disable
dynamism. One cursor per phase is cleared at the token boundary, avoiding a
second grid barrier to reset the cursor before every phase.

All workers reach phase completion, including workers with no useful tile.
Consequently, the design trades explicit task dependencies and per-task queue
traffic for a grid-wide barrier. Uniform projection tiles can favor cyclic
assignment, whereas unequal attention service times may benefit from claims.
We retain both policies and measure their cost rather than defining dynamic
scheduling to mean that atomic stealing must always win.

\subsection{Fused Tensor Core Handlers}
Each of Qwen's 28 layers executes six phases: normalization plus Q/K/V
projection; head normalization, RoPE and KV publication; attention; output
projection plus residual; normalization plus gate/up projections and
SiLU--multiply; and down projection plus residual. Final normalization,
vocabulary projection and greedy selection complete the token.

Projection handlers use BF16 Tensor Core operands and FP32 accumulators.
A tile reuses weight loads across up to four live requests. The gate and up
projections execute in paired warps for the same output rows, then combine
their results in shared memory before writing the MLP activation. BF16
rounding at the original projection and activation boundaries is retained.
This combines horizontal gate/up work with vertical activation fusion: it
reduces the number of projection claims, eliminates the materialized gate/up
tensor, and removes one phase per layer. The separate-MLP control uses the same
projection arithmetic but writes gate/up results and launches SiLU separately.

Attention processes physical 128-key tiles with masked online softmax, BF16
probability operands and FP32 accumulation. Physical KV capacity is rounded
to a multiple of 128 for both native and custom paths; logical contexts and
output limits do not change. A sequential projection accumulation and matched
normalization reduction tree improve compatibility, but identical data types
alone do not guarantee native-framework equivalence. Section~\ref{sec:numerics}
reports the unchanged full-model numerical gate.

\subsection{Progress and Ordering}
Atomic claims return distinct indices; cyclic indices are also injective.
Each tile owns disjoint output rows or request/head pairs. Only producer-ready
work is admitted, so a waiting consumer cannot occupy a worker needed by an
unscheduled producer. Bounded handlers finish before the collective barrier,
whose completion makes their writes visible to the next phase.

Before launch, \sys checks cooperative support and the occupancy limit for
the actual compiled kernel and dynamic shared allocation. The worker grid must
fit simultaneous residency. Induction over phases establishes dependency
ordering; induction over layers and bounded token iterations establishes
termination under valid pointers, supported dimensions and successful device
execution. The argument depends on every handler participating in its required
barriers and completing asynchronous operations before release. It is not a
contract for arbitrary blocking user code.

Compaction writes to an earlier or equal map position, preserving unread IDs
in a forward scan. A barrier separates the update from all readers. Already
published outputs remain indexed by stable ID after a request retires.
This ownership rule and the last-consumer rule below jointly prevent a live
request from observing another request's recycled intermediate.

\section{Allocation Through Phase Lifetimes}
\label{sec:workspace}
The key allocation decision is the lifetime of a region, not the frequency of
calling an allocator.\aidraft{} \sys uses phase completion to release two
different resources: block-local scratch for the next handler, and global
intermediate regions for the next consumer chain. Their physical reservations
and their logical usage are accounted for separately.

\subsection{Block-Local Shared Storage}
A persistent kernel retains its allocated shared memory across all handlers.
Independent static arrays in the paired and unpaired projection specializations
can therefore accumulate even though the handlers never overlap. Our
counterfactual keeps those separate arrays, while the main implementation maps
their normalized inputs and partial accumulators onto one 12-KiB dynamic region.
Attention uses the same region in its own phase.

Within a handler, block barriers protect producer/consumer exchanges. At the
last consumer, the phase barrier establishes that no worker can still access
the preceding handler's scratch. Reinterpreting the region for the next handler
is then safe. This is direct lifetime overlay, with neither shared-memory
free-list operations nor a promise of cross-phase pipelining.

The complete block allocation, including remaining static state, decreases
from 36,704 to 21,856 bytes. The occupancy API limit changes from two to three
blocks per SM, but the main experiment launches the same 340 blocks on 170 SMs
in both builds. Thus a higher theoretical occupancy limit alone cannot explain
the measured improvement. Compiler registers, actual activity and memory
traffic are examined separately; storage fusion changes more than one resource.

\subsection{Global Region Plan}
For $b$ live requests, the fused decoder needs two BF16 regions,
\begin{equation}
R_{\mathrm{qkv}}=4096b,\qquad R_{\mathrm{attn}}=2048b
\end{equation}
elements, totaling $12b$ KiB. The Q/K/V region becomes reusable after attention
and the output projection consume it. The fused MLP then stores $3072b$
activation elements in that region; the down projection completes before the
next layer overwrites it. All 28 layers reuse the same regions. The separate
MLP additionally materializes $6144b$ gate/up elements, raising the plan to
$24b$ KiB while preserving the same cross-layer reuse.

A checked monotone bump assigns region offsets and rejects any extent beyond
the host-reserved arena. Metadata records the high-water mark, region count
and exhaustion flag. An exhausted invocation terminates collectively before
publishing a decode token for the failed step. Guards check that no write
escapes the reservation. Weights, KV, logits, residual state and framework
prefill buffers are outside this transient-scratch accounting.

\subsection{Fixed Capacity Versus Dynamic Rebuilding}
The active-set implementation rebuilds offsets at each decode step. A decisive
counterfactual constructs one plan at the admitted batch capacity and keeps
those offsets for the invocation. Both versions compact exactly the same
requests, execute exactly the same live tiles, and reuse regions across every
layer. Thus this comparison changes planning work, not padded computation or
the opportunity for cross-layer reuse.

The fixed plan has the same capacity and high-water mark. In the four-request
ragged case, rebuilding performs 62 region assignments over 31 decode steps,
whereas the fixed plan performs two. Their measured latencies are nearly equal.
This result narrows the mechanism: phase-lifetime reuse and avoiding materialized
intermediates matter; rebuilding a bounded region plan is neither necessary
for request retirement nor a demonstrated speed advantage. A future admission
policy could use the released logical capacity, but this prototype does not
return it to a global allocator or admit new requests during decode.

\subsection{When a Fixed Plan Is Sufficient}
The counterfactual suggests a useful implementation rule. Suppose a phase
schedule has a fixed set of alias groups, a maximum admitted batch $B$, and a
known bound $S_j(b)$ on the storage needed by group $j$ at live count $b$.
Reserve fixed, aligned regions with capacity
\begin{equation}
C=\sum_j \max_{0\leq b\leq B}\operatorname{align}(S_j(b)).
\end{equation}
If all handlers address a region through the same group base and reuse it only
after its last consumer, request retirement needs to change valid extents and
task bounds, not region bases. For this decoder, the two alias groups have
linear sizes, so the maxima occur at $B$ and $C=12B$ KiB.

To see the equivalence, assume both plans start a phase with the same values
for every live request. Stable IDs select the same persistent inputs, and
packed positions select corresponding intermediate values even when absolute
region offsets differ. Disjoint tile ownership produces the same next-phase
values. Last-consumer completion makes the reused region dead before its next
write. Induction over phases and tokens therefore preserves outputs and
retirement decisions; the capacity bound prevents an address from escaping its
reservation. This argument requires the same handler arithmetic and a common
bounded lifetime schedule, rather than arbitrary asynchronous allocation.

This separates two kinds of adaptation. Work membership and valid tensor
extents respond to actual model outputs. Physical capacity and alias-group
bases can remain fixed until the invocation returns. Replanning becomes
necessary when admission exceeds the reserved bound or the dependency/lifetime
structure changes, not merely because a request finishes. The fixed-plan
experiment tests this distinction without confounding it with padded work or
removing cross-layer reuse.

\section{Evaluation Method}
\label{sec:method}
The evaluation asks whether fusion and lifetime reuse improve a dynamic
persistent executor, whether the benefit survives strong launch controls, and
what changes when requests share a software pool instead of separate MPS
clients.\aidraft{} Every comparison identifies its arithmetic and timing
boundary; numerical compatibility is an independent acceptance gate.

\subsection{Platform and Timing}
Experiments use an RTX 5090 with 170 SMs, compute capability 12.0 and 32 GB
memory, CUDA 13.1 targeting \code{sm_120}, and driver 610.57.04. The host is an
AMD Threadripper 7960X with 24 cores and 48 threads. The GPU's maximum PCIe
link is Gen5 $\times16$, attached to NUMA node 0. Main processes use logical
CPU 2; its SMT sibling 26 is not isolated. MPS clients use CPUs 2/4/6/8.
The post-reboot environment records the powersave governor with boost enabled;
CPU and GPU clocks are unlocked and the display remains active. GPU compute
process checks reject unrelated jobs before primary cells, at MPS round startup
and after client connections. A reboot separates this suite from historical experiments; their
timings are never pooled.

The environment pins Python 3.14.4, PyTorch 2.14.1+cu130, Torchvision
0.29.1+cu130 and Transformers 4.57.6. TF32 and BF16 reduced-precision GEMM
reductions are disabled; BF16 products accumulate in FP32. CPU tensor execution
uses one thread and tokenizer parallelism is disabled. The host monotonic
\code{perf_counter_ns} timer brackets explicit CUDA completion synchronization.
Model loading, compilation, capture, allocation and warmup are outside warmed
request time. Caches are warmed rather than flushed.

\subsection{Model Inputs and Full Request Boundary}
We use the official Qwen3-0.6B BF16 weights~\cite{QwenModel}, revision
\code{c1899de289a04d12100db370d81485cdf75e47ca}. Single uses prompt/output
lengths 256/64; Uniform has four 32/32 requests. Ragged uses contexts
16/32/64/128 and limits 8/16/24/32. Long uses contexts 129/257/385/513 and limits
9/17/33/65. Skew-8 uses contexts 8/16/24/40/80/128/192/257 and limits
2/3/5/8/13/21/34/55. These authored prompts are repeated and suffix-cropped to
the specified contexts. They exercise actual model operators, but are not a
production trace or a language-quality dataset.

EOS uses four retained short-answer instructions at contexts 64/128/192/256,
with actual tokenizer EOS and a maximum of 96 tokens. Its stopping times are
model outputs, not a host-specified retirement schedule. Other cases disable
EOS to hold the amount of output fixed. Thinking is disabled and decoding is
greedy. The native runner must reproduce Hugging Face generation on every
required reference token.

A complete local request includes CPU tokenization/padding, H2D, native prefill,
all required decode steps, and D2H/text decoding. Decode time begins after the
first prefilled token has been retrieved. Native graphs capture prefill and
one decode step; custom controls and \sys retain the same native eager prefill.
These differences remain inside complete-request totals. EOS native graph
execution polls completion between token replays; the custom controls evaluate
the loop on the GPU. Network transport and admission queueing are outside the
request boundary.

\subsection{Matched Fusion and Launch Controls}
The primary factorial comparison crosses fused/separate MLP handlers with
persistent, tile-grid+PDL, and CLC+PDL execution. The two ordinary-kernel controls
use a GPU conditional WHILE loop, the same active-map update and the same global
arena policy. The tile grid exposes one logical block per task, with no
340-block budget. CLC cancels unstarted block indices using CUDA 13.1's CCCL
implementation. Both use programmatic edges: each successor acquires its
predecessor dependency before any dependent read. Their handlers provide
little independent prologue work, limiting the available PDL overlap.

Persistent runs use 340 blocks of 128 threads. Additional controls remove
shared-storage overlay, build one fixed region plan, replace atomic claims
with cyclic assignment, or remove PDL from the ordinary tile grid. All matched
paths must reproduce fused-persistent tokens and final logits bitwise.
Source and binary hashes are frozen before three independently started process
rounds. Cell order is randomized; inputs and input seed remain fixed. Each
cell has three warmups and ten timed requests. The complete matrix has 138
cells and 1,380 requests. Values are medians of three process medians; intervals
span those medians and are not confidence intervals. Paired ratios are computed
within a process before aggregation.

\subsection{Real MPS Sharing Comparison}
Four spawned CUDA processes connect to a private legacy-v2 MPS server. The
retained server client list verifies all four connections. Clients share model
weights through CUDA IPC and own private KV and scratch. Each receives its row
from the \emph{same native pooled-prefill checkpoint}; the RMS reduction tree
is fixed to the pooled batch capacity. Tokens and logits must match the pooled
custom result bitwise, including EOS retirement. This prevents different
prefill arithmetic or replicated weights from explaining the sharing result.

We compare pooled persistent, pooled CLC+PDL, four MPS tile+PDL clients, four
MPS CLC+PDL clients, and CLC+PDL clients capped at 25\% active threads each.
The latter is a soft MPS provisioning setting, not MIG emulation. Every trial
restores the checkpoint outside timing and synchronizes all clients before a
common release. Group makespan ends after all four outputs reach CPU text.
The MPS interval includes release-barrier overhead; process startup, checkpoint
copy, prefill and IPC result collection are excluded. This is a post-prefill
sharing experiment, separate from the complete-request benchmark.

Both Ragged and EOS use three fresh process rounds, randomized method order,
three warmups and ten timed groups per method: 300 groups in total. The private
MPS daemon remains available during a round; pooled paths use the parent CUDA
context created before that namespace. Raw long samples and process ranges
are retained without latency-based filtering.

\subsection{Numerical and Mechanism Checks}
Native compatibility requires finite same-prefix final logits, normalized RMSE
at most 1\%, and cosine similarity at least 0.999. The threshold is fixed before
the runs; free-running token agreement is reported separately. Failed cases
remain in the custom-control analysis but cannot establish an equivalent-native
speedup. Every timed request validates its token hash and each custom cell
checks its arena guards.

Selected three-step, four-request decodes are profiled with Nsight Compute.
Profiler timing is excluded from performance aggregates; counters provide
selected-point evidence about activity, traffic and stalls. An isolated empty
grid-barrier benchmark supplies a lower bound rather than an additive estimate
of full decode cost. Functional tests cover odd contexts, unequal retirement,
fixed/rebuilt regions, exhausted arenas and rejected oversized cooperative
grids. The new fused kernel is also checked with memcheck and synccheck.

\section{Evaluation of Fused Dynamic Execution}
\label{sec:applications}
We first isolate fusion and execution, then test allocation and assignment,
complete requests, and actual process sharing.\aidraft{} The same-handler
controls answer mechanism questions even when native numerical compatibility
fails; the latter is reported explicitly in Section~\ref{sec:numerics}.

\subsection{Fusion Changes the Execution Ranking}

Figure~\ref{fig:factorial} puts the strongest matched launch controls beside
the persistent executor. Fused persistent decode takes 75.7, 78.9, 254.0 and
210.2 ms on Uniform, Ragged, EOS and Long. Separate-MLP persistence takes
96.4, 101.6, 327.4 and 263.5 ms. The benefit of the combined gate/up/activation
handler is therefore visible within an unchanged persistent scheduling policy.
It removes a phase and a materialized intermediate, halves the gate/up task
count for a four-request group, and retains the same BF16 rounding boundaries.

Fusion alone is insufficient. Under CLC+PDL, the fused path takes 88.1, 98.3,
334.4 and 287.2 ms, while its separate-MLP counterpart takes 86.1, 92.6,
303.6 and 252.6 ms. The more fused variant loses in all four cases. Under
tile-grid+PDL, fusion helps, but the resulting times remain 86.6, 93.0, 305.7
and 253.8 ms. The winning decomposition depends on the execution mechanism;
a smaller phase count is not a sufficient optimization criterion.

Allowing each nonpersistent control its faster MLP variant, the within-process
ratio to fused persistence has median 1.139, 1.173, 1.194 and 1.198 across the
four cases. This comparator takes the minimum among measured tile+PDL and
CLC+PDL variants in each process, rather than selecting a weak fused baseline.
The two additional shapes compare fused persistence with fused CLC+PDL: Single
takes 168.5 versus 224.2 ms and Skew-8 takes 155.1 versus 198.3 ms. Those cases
do not contain the complete factorial matrix, so their ratios are not presented
as a search over all decompositions.

The separate tile-grid control further identifies PDL's contribution. On
Uniform and Ragged, fused ordinary grids take 88.7 and 95.1 ms; adding PDL
reduces them to 86.6 and 93.0 ms. The resulting gain is smaller than the
remaining difference to persistent execution for these dependent phases.

\subsection{Storage, Planning and Claiming Counterfactuals}
\begin{figure*}[t]
\centering\includegraphics[width=\textwidth]{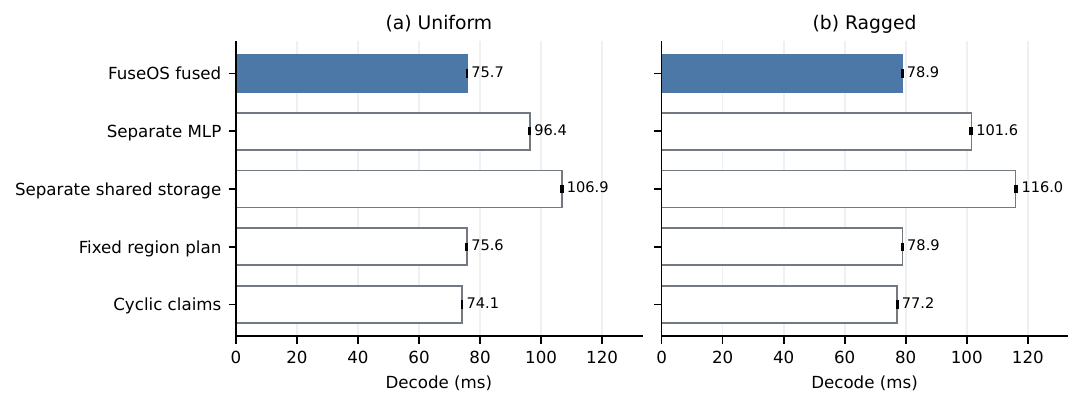}
\caption{Complete-decode ablations with identical outputs. Separate shared
storage preserves the fused computation but retains independent static arrays
for projection specializations. Fixed planning preserves compaction and
cross-layer reuse. Cyclic assignment preserves dynamic work bounds. Bars and
intervals use the same process-median convention as Fig.~\ref{fig:factorial}.}
\label{fig:ablation}
\end{figure*}

Separate shared storage increases Uniform/Ragged decode to 106.9/116.0 ms,
versus 75.7/78.9 ms with overlay (Fig.~\ref{fig:ablation}). It is even slower
than the separate-MLP persistent implementation. Thus simply fusing the
operator code, without controlling the resulting storage lifetime, can erase
the measured advantage. The source change also changes compiler register
allocation, so this comparison is a storage-design ablation rather than a
claim that byte count alone explains every cycle.

Fixed global planning takes 75.6/78.9 ms, effectively matching per-token
rebuilding. Both reserve 48 KiB for four requests and use the same last-consumer
reuse across 28 layers. Rebuilding has no measured performance advantage in
this contract. Cyclic claims take 74.1/77.2 ms, and are also modestly faster on
EOS and Long (247.9/204.6 versus 254.0/210.2 ms). These dense tiles favor avoiding
atomic assignment overhead. The main atomic configuration remains fixed
throughout the comparisons; dynamism does not depend on selecting it.

\begin{figure*}[t]
\centering\includegraphics[width=\textwidth]{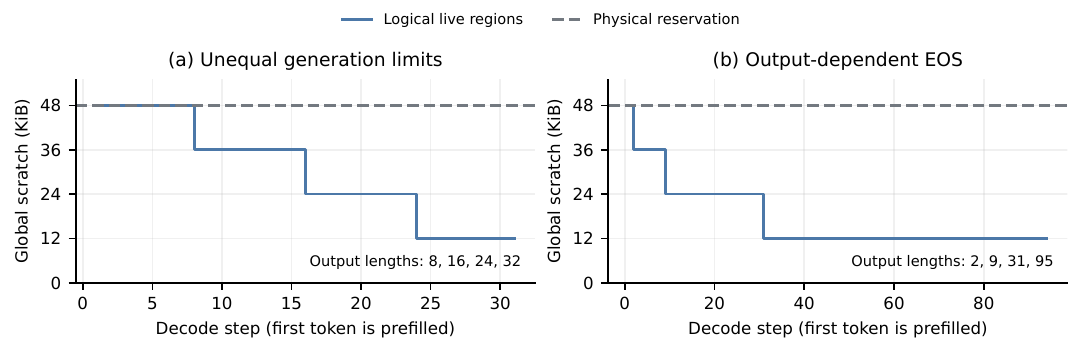}
\caption{Logical scratch reconstructed from verified output lengths. Ragged
uses supplied generation limits; EOS lengths 2/9/31/95 are produced by the
custom model. The host reservation stays at 48 KiB. These are reconstructions
of the checked region contract, not sampled device-memory utilization traces.}
\label{fig:lifetimes}
\end{figure*}

Figure~\ref{fig:lifetimes} shows what retirement actually changes. The logical
global plan decreases from 48 to 12 KiB as requests leave, while the physical
reservation stays fixed. Projection task counts change in groups of four live
requests, so retiring a single request need not reduce the number of Tensor
Core tiles. It still removes its normalization, valid output and attention
work. Skew-8 crosses this grouping boundary; EOS exercises output-dependent
retirement without a precomputed host schedule.

\subsection{Testing the Resource Explanation}
\begin{table}[t]
\caption{Selected-point resource evidence: four requests, contexts
16/32/64/128, three remaining decode steps, 340 blocks. F: fused overlay;
S: separate MLP with overlay; A: fused with separate static arrays.
Counters are from profiling, not the latency aggregate.}
\label{tab:resources}\centering\setlength{\tabcolsep}{4pt}
\begin{tabular}{@{}lrrr@{}}\toprule
Quantity & F & S & A \\\midrule
Registers/thread & 168 & 130 & 134 \\
Shared bytes/block & 21,856 & 21,856 & 36,704 \\
Occupancy API blocks/SM & 3 & 3 & 2 \\
Measured active warps (\%) & 16.67 & 16.66 & 16.66 \\
DRAM read (GB) & 3.96 & 3.95 & 3.95 \\
L2 requested traffic (GB) & 5.40 & 5.63 & 5.58 \\
CTA barrier stall (\%) & 63.69 & 64.72 & 65.50 \\
\bottomrule\end{tabular}
\end{table}

Table~\ref{tab:resources} rejects two simple explanations for the overlay
benefit. Achieved active-warps fractions are almost identical because all
variants use the same grid, and DRAM read traffic is approximately unchanged.
The compiler uses different register counts, while L2 requested traffic is
lower for fused overlay. A second profile finds 442.1 versus 463.9 million
issued warp instructions for overlay versus separate arrays, at similar
measured SM frequencies (2.91 versus 2.92 GHz). Shared-load bank conflicts and
local-memory sector counts are close. These observations identify differences
in generated execution without assigning the entire gain to one counter.

We then hold the \emph{device instructions} fixed and change only launch-time
dynamic shared reservation from 12,288 to 27,136 bytes. The larger reservation
matches the separate-array build's total 36,704-byte block footprint.
The complete decoder SASS is identical, verified byte-for-byte after removing
object-file metadata. Three process rounds, with three warmups and ten trials
per setting, retain identical outputs. Uniform takes 75.60 versus 75.59 ms;
Ragged takes 78.75 versus 78.75 ms. Thus the slower separate-array variant
cannot be explained by shared reservation size alone at this worker budget.
The demonstrated gain belongs to the lifetime-overlay implementation and its
generated code; a pure occupancy explanation would contradict this control.

The remaining costs are also distinct. Cyclic assignment measures the effect
of removing claims, and tile grids with/without PDL measure a launch-policy
change. Separate empty-grid experiments report roughly 0.6--0.8~$\mu$s per
cooperative barrier. The profiler's CTA-barrier stall fraction refers to
intra-block waits, not an additive fraction of grid-synchronization latency.
Fusion changes task grouping, activation materialization and synchronization
together. Adding these independently measured costs would not produce a valid
decomposition of complete decode time.

\subsection{Native Compatibility and Complete Requests}
\label{sec:numerics}
\begin{table}[t]
\caption{Native compatibility and complete local-request time. Medians of three process medians; NRMSE uses the same reference prefix, while token agreement uses free greedy generation. NG: native graph; F: fused persistent.}
\label{tab:native}\centering\setlength{\tabcolsep}{3.8pt}
\begin{tabular}{@{}lrrrrr@{}}\toprule
Case & NG (ms) & F (ms) & NRMSE\% & Tokens\% & Gate \\ \midrule
Single & 235.7 & 184.1 & 0.93 & 100.0 & pass \\
Uniform & 131.8 & 92.1 & 1.07 & 84.4 & fail \\
Ragged & 145.8 & 96.0 & 1.01 & 93.8 & fail \\
EOS & 455.2 & 275.3 & 1.18 & 58.2 & fail \\
Long & 366.8 & 239.4 & 0.96 & 95.2 & pass \\
Skew-8 & 315.2 & 186.4 & 1.19 & 73.8 & fail \\
\bottomrule\end{tabular}
\par\smallskip\footnotesize Fixed gate: finite logits, NRMSE $\leq1\%$, cosine $\geq0.999$. A failed gate excludes an accuracy-equivalent native speedup. Long passes the logit gate but does not reproduce every greedy token.
\end{table}

\begin{figure*}[t]
\centering\includegraphics[width=\textwidth]{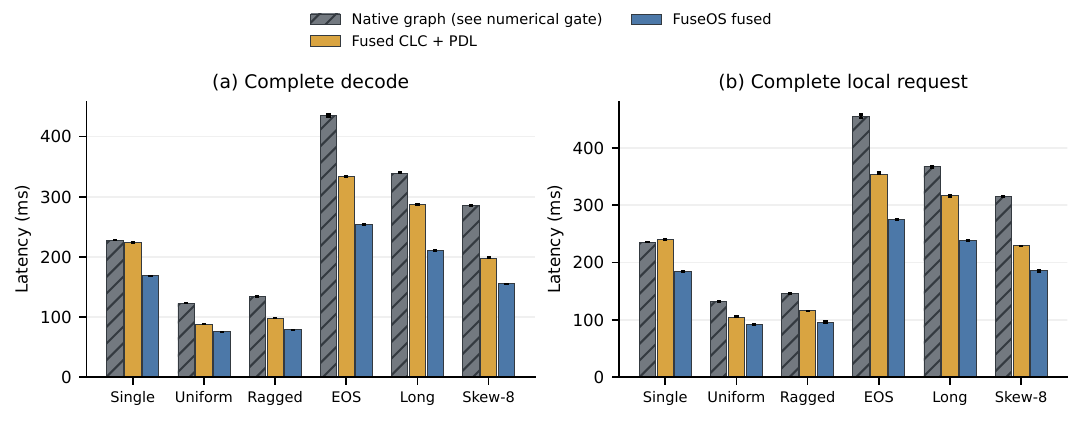}
\caption{Complete decode and local-request boundaries for Qwen3-0.6B.
Request time includes CPU input/output and prefill. Native graph uses different
operator arithmetic; its comparison is subject to Table~\ref{tab:native}.
Failed native gates are retained as timings, not claimed as accuracy-equivalent
application speedups. Intervals span three process medians.}
\label{fig:requests}
\end{figure*}

The numerical investigation identified a concrete arithmetic mismatch: the
framework's default permits reduced-precision BF16 GEMM reductions, whereas
the custom projections accumulate in FP32. Disabling that reduction for all
new paths reduced operator-local down-projection/residual disagreements from
hundreds of BF16 elements to zero or one in retained diagnostic cases. We also
match normalization reduction structure and masked online attention tiling.
These changes improve a known compatibility issue without relaxing the gate.

The remaining full-model result is mixed. Single and Long pass at 0.934\%
and 0.965\% NRMSE; Uniform, Ragged, EOS and Skew-8 fail at
1.073\%, 1.011\%, 1.181\% and 1.191\%. All native reference tokens agree with
Hugging Face generation. Free-running custom agreement ranges from 58.2\%
for EOS to 100\% for Single. EOS native lengths are 2/7/31/94, whereas custom
lengths are 2/9/31/95. Accordingly, the native EOS timing does not even represent
exactly the same generated workload. Every matched custom control and MPS
client does reproduce the custom workload bitwise.

Single provides the strongest qualified application comparison: its complete
request takes 184.1 ms with \sys and 235.7 ms with native graph, with all greedy
tokens matching. Figure~\ref{fig:requests} also shows how common prefill and CPU
work reduce gains seen in decode alone. Long passes the logit gate but changes
some greedy tokens, so passing that gate is not presented as exact generation
equivalence. Broader model-quality validation and the four failed numerical
cases remain necessary before deployment as a drop-in native replacement.

\subsection{Pooling Versus Four MPS Clients}
\label{sec:sharing}
\begin{figure*}[t]
\centering\includegraphics[width=\textwidth]{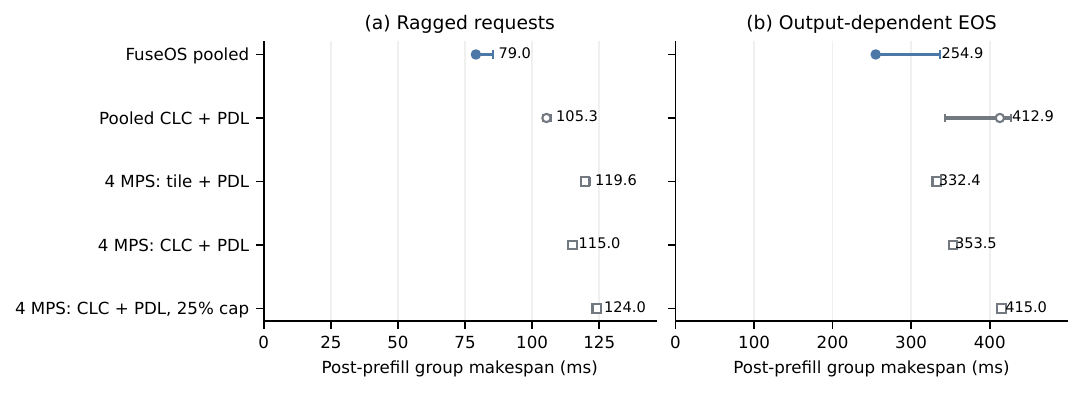}
\caption{Real MPS comparison from identical prefill checkpoints, with shared
weights and bitwise-identical custom outputs. The metric is post-prefill
four-request group makespan through D2H/text, not full request time. Intervals
retain all three process medians. The 25\% setting is an active-thread cap;
clients report 42 visible SMs instead of 170. It is not a MIG partition.}
\label{fig:mps}
\end{figure*}

On Ragged, pooled persistence takes 79.0 ms, pooled CLC+PDL 105.3 ms,
four MPS CLC+PDL clients 115.0 ms, and four MPS tile+PDL clients 119.6 ms.
The paired MPS-CLC/persistent ratio is 1.455. Pooling alone supplies part of
the advantage, while the pooled CLC control shows an additional benefit from
persistent execution. Shared weights and identical prefill checkpoints rule
out weight replication and different output work as explanations.

On EOS, pooled persistence has a 254.9-ms median, compared with 332.4 ms for
MPS tile+PDL and 353.5 ms for MPS CLC+PDL. The paired MPS-tile/persistent ratio
has median 1.307, but one persistent process median reaches 336.8 ms and is
slower than its 331.3-ms MPS-tile counterpart. Pooled CLC+PDL is also variable,
with process medians 343.2--426.5 ms, and loses to MPS tile+PDL. Pooling is
therefore not sufficient to win independently of executor choice, nor does
the median establish a latency bound.

The 25\% capped CLC clients take 124.0 ms on Ragged and 415.0 ms on EOS,
slower than unrestricted MPS. Released logical work does not guarantee that
a per-client cap exposes the remaining request's full parallelism. The evidence
supports a software-pool alternative for compatible trusted requests on this
GPU. It does not measure sustained arrivals, fairness across heterogeneous
models, or replacement of hardware isolation. The benchmark's finite admitted
groups and observed run variation define the demonstrated sharing boundary.

\section{Integration Boundaries}
\label{sec:discussion}
\sys's useful execution region contains known handler types, runtime-dependent
work extents, and dependencies expressible as phases.\aidraft{} Its evidence
supports coupling that region to one request pool and explicit storage
lifetimes. The two integrations below clarify why coverage of that region
matters more than adding persistence to an isolated operator.

\subsection{ResNet-18: Partial Persistence Can Lose}
\begin{figure*}[t]
\centering\includegraphics[width=\textwidth]{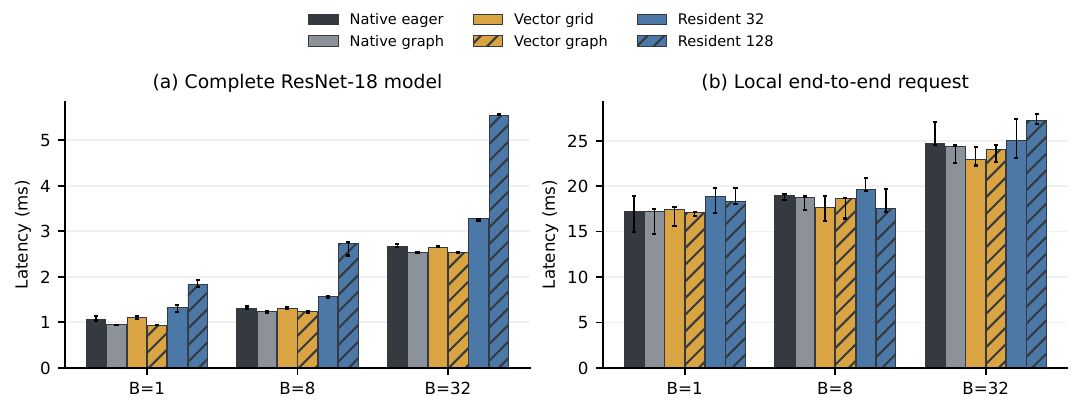}
\caption{Retained complete ResNet-18 boundary experiment, separate from the
new Qwen suite. All paths use folded batch normalization, cuDNN autotuning and
NCHW layout. Vector and resident paths share the same float4 residual handler
and reused arena. Three process medians of thirty requests determine bars and
ranges. Request time includes one photograph's preprocessing and batch
replication; it is not independent-image serving throughput.}
\label{fig:resnet}
\end{figure*}

The ResNet adapter replaces eight residual add--ReLU operations with resident
handlers while leaving convolution, pooling and classification in the framework.
A stream-ordered publication follows each producer convolution; the consumer
stream waits for the residual handler before reusing its buffer. Inference-mode
last use permits eight residual outputs to share a region sized for the largest
tensor. Ordinary vector-grid and vector-graph controls receive the same handler,
batch-normalization folding and workspace reuse.

This retained suite uses pretrained ImageNet-1K V1 ResNet-18
weights~\cite{TorchvisionResNet}, FP32, and batches 1/8/32 at $224\times224$.
One retained JPEG is decoded, transformed and replicated. Three processes each
measure thirty requests after three warmups; the earlier host environment uses
the performance governor. These data are not pooled with post-reboot Qwen
timings. All cells pass the numerical and top-1 checks against the original
unfolded model.

At batch 32, native graph model time is 2.534 ms and vector graph time is
2.536 ms; resident adapters with 32/128 workers take 3.273/5.553 ms.
The native graph also wins model time at smaller batches. Request time is
dominated by CPU image handling: the native graph batch-32 request takes
24.383 ms despite 2.534-ms model execution. These measurements demonstrate a
coverage boundary. Persistent residual execution retains eight publication and
completion exchanges, while the framework still submits the expensive
convolutions. Buffer reuse removes allocations, but does not remove those
coordination boundaries. Qwen instead keeps the complete decode dependency
region and token loop inside the executor.

\subsection{Optional TMA Is a Separate Handler Optimization}
An earlier elementwise suite evaluates one/two-stage bulk and tensor TMA with
saved barrier parity and last-consumer reuse~\cite{CUDATMA,CUDATensorMap}.
At 32 workers, a large-task float4 path takes 484.3~$\mu$s and two-stage tensor
TMA takes 319.0~$\mu$s. An independently matched one-round ordinary subdivided
grid takes 105.0~$\mu$s versus 111.3~$\mu$s for 128-worker TMA. Thus a
worker-local staging benefit does not justify restricting available parallelism.
TMA is not integrated into the measured Qwen path and supplies no multiplier
for its inference results. Its implementation, original figures and raw
controls remain in the artifact rather than occupying the main evaluation.

\subsection{Scope of the System Claim}
The implemented allocation abstraction is a checked transient arena and
block-local overlay. It does not manage arbitrary device allocation lifetimes,
free KV pages for new clients, or guarantee tenant isolation. A pooled
execution domain is an alternative to separate-client placement for trusted
requests; MPS and MIG remain appropriate when their sharing or isolation
contracts are required. No MIG performance value is inferred from the MPS cap.

The benchmark admits finite groups, not a continuous service. It measures one
GPU, fixed authored inputs, unlocked clocks and three process rounds. In
particular, the observed EOS variation prevents turning a median benefit into
a tail-latency guarantee. The prototype is specialized to Qwen3-0.6B and does
not establish downstream language accuracy. Four native numerical failures
and missing empirical MPK/Meld comparisons remain material limits on generality.

\section{Conclusion}
\sys couples dynamic request membership, fused tile execution and storage
lifetimes inside a persistent GPU resource pool.\aidraft{}
The complete decode integration outperforms the strongest measured matched
launch controls in the factorial cases, while fixed-plan and cyclic controls
identify mechanisms that need not be dynamic on every token. Real MPS tests
show the opportunity and variability of moving sharing from client boundaries
to a common ready-work pool. The resulting design principle is concrete:
choose fusion, execution and storage lifetimes together, then test them against
the same arithmetic and live-work contract.


\bibliographystyle{IEEEtran}
\bibliography{references}
\end{document}